\documentclass[preprint]{jpsj3}

\usepackage{graphicx}
\usepackage{color}

\title{
Weak-coupling Mean-field Theory of Magnetic Properties of NiGa$_2$S$_4$ 
}

\author{
Takuji Nomura$^{1,2}$\thanks{E-mail address: nomurat@spring8.or.jp}, 
Yuji Yamamoto$^2$, Kenji Yoshii$^3$}

\inst{
$^1$ Synchrotron Radiation Research Center, 
National Institutes for Quantum and Radiological Science and Technology, 
SPring-8, 1-1-1 Kouto, Sayo, Hyogo 679-5148, Japan \\
$^2$ Graduate School of Material Science, 
University of Hyogo, 3-2-1 Kouto, Kamigori, 
Hyogo 678-1297, Japan \\
$^3$ Materials Sciences Research Center, Japan Atomic Energy Agency, 
SPring-8, 1-1-1 Kouto, Sayo, Hyogo 679-5148, Japan \\}

\recdate{\today}

\abst{
We report a mean-field theoretical study of a triangular lattice magnet NiGa$_2$S$_4$. 
Specifically, spiral mean-field theory is applied to a 17-band $d$-$p$ model 
constructed from the maximally localized Wannier functions. 
Our itinerant-model approach shows that the most stable spiral magnetic state 
has an ordering vector near ${\mib Q}=(0.15,0.15,0)$, consistent with neutron scattering experiments, 
when we assume the Ni-site Coulomb interaction is not so strong ($U \approx 2$ eV). 
To map onto a classical Heisenberg spin model, we estimate spin exchange interactions from the mean-field results, 
and find that the nearest-neighbor exchange is ferromagnetic and the largest 
(larger than the third nearest-neighbor exchange, in contrast to early studies). 
We also calculate the dynamical spin correlation function $S({\mib q}, \omega)$, 
using the same model within the random-phase approximation (RPA). 
Calculated $S({\mib q}, \omega)$ has a spectral structure quite different 
from that of conventional spin-wave excitations.} 

\begin{document}
\maketitle

\section{Introduction}
\label{Sc:1}

Magnetism on a regular triangular lattice has been studied with considerable interest, 
because such a lattice structure with geometrical frustration generally 
prohibits conventional collinear spin configurations 
and may realize exotic spin states when nearest-neighbor spins are coupled antiferromagnetically. 
Among regular triangular lattice magnets, NiGa$_2$S$_4$ has been considered as a possible realization 
of novel exotic magnetic state~\cite{Nakatsuji2005, Nakatsuji2010}. 
The crystal of NiGa$_2$S$_4$ is constructed by stacking slabs along the $c$-axis, 
in each of which a NiS$_2$ layer is sandwiched by a pair of GaS layers. 
The nearest-neighboring three Ni sites on a NiS$_2$ layer form a regular triangle parallel to the $ab$ plane. 
Since the distance between nearest NiS$_2$ layers exceeds three times of that between in-plane nearest-neighbor Ni atoms, 
its magnetic correlations are naturally expected to be of strong two-dimensionality. 
In fact neutron scattering suggests only weak interlayer ferromagnetic correlation~\cite{Stock2010}. 
Concerning the in-plane magnetic correlation, neutron scattering reveals clearly 
the incommensurate antiferromagnetic correlation with the in-plane propagation vector 
${\mib Q} \equiv (0.15(5), 0.15(5))$~\cite{Stock2010}, 
which differs from the 120-degree ordering (${\mib Q}=(1/3, 1/3)$) conventionally expected for antiferromagnetic Heisenberg models. 
This incommensurate correlation has been explained by assuming nearest-neighbor ferromagnetic spin coupling $J_1$ 
and larger third nearest-neighbor antiferromagnetic $J_3$ with $ J_1/J_3 \approx -0.2$~\cite{Nakatsuji2005}. 
One of the most remarkable features is that the in-plane magnetic correlation length, 
which is estimated from the reciprocal of the Bragg peak width at ${\mib Q}$, 
remains only up to about seven times of the nearest-neighbor Ni distance without showing divergence, 
i.e., the long-range ordering is not reached, even at 25 mK~\cite{Nakatsuji2005, Nakatsuji2010, Stock2010, Nambu2015}. 

Unusual magnetic properties of NiGa$_2$S$_4$ have been characterized by various measurements, 
e.g., NMR-NQR~\cite{Takeya2008}, $\mu$SR~\cite{Takeya2008,Yaouanc2008,MacLaughlin2008}, ESR~\cite{Yamaguchi2010}. 
First, the bulk DC susceptibility shows a clear kink at $T^*$~\cite{Nakatsuji2005,Nakatsuji2010}, 
which reminds us of apparently antiferromagnetic or spin-glass transition at $T^*$. 
The Ga-NQR relaxation rate is critically enhanced, being too large to observe, 
in a wide temperature range $T=$2 K - 10 K around $T^*$. 
This indicates that there exists slow spin dynamics with the spins not freezing immediately 
below $T^*$ but keeping fluctuations down to 2 K. 
Such persistent spin dynamics below $T^*$ clearly distinguishes NiGa$_2$S$_4$ from conventional antiferromagnets 
in which dynamical spin motions are quenched abruptly or rapidly below the transition temperature. 
Below 2 K, Ga-NMR and NQR spectra become extremely broad, indicating inhomogeneous or incommensurate 
static magnetic ordering with frozen spins~\cite{Takeya2008}. 
Also $\mu$SR, which is capable of observing slower spin dynamics than NQR, detects the occurrence 
of internal inhomogeneous magnetic fields with a mean-field-like behavior below $T^*$~\cite{MacLaughlin2008}. 
Spin freezing sets in below $T^*$, but spin relaxation persists below $T^*$ down to 2 K, 
basically agreeing with the Ga-NMR and NQR measurements. 

By summarizing the above observations in a consistent way, 
we are led to the following view : 
some magnetic transition occurs at $T^*$, accompanied by spontaneous spin polarization at Ni site, 
but the polarized spins keep fluctuation persistently down to 2 K. 
At the lowest temperature below 2 K, static incommensurate magnetic ordering with frozen spins occurs, 
whose spatial spin configuration is characterized by the wavevector ${\mib Q}$, 
but long-range ordering signaled by the divergence of the correlation length 
is not reached even at the lowest temperature. 
The occurrence of internal fields due to the spontaneous spin polarization excludes 
the realization of singlet spin liquid state at the low temperatures. 

Concerning the thermodynamic property, the magnetic part of specific heat shows $C_M \propto T^2$ behavior below $T^*$ , 
suggesting existence of some linearly dispersive modes at low energy~\cite{Nakatsuji2005, Nakatsuji2010}. 
However, the specific heat exhibits no visible anomaly at $T^*$, 
while it has broad humps around 10 K and 80 K. 
Furthermore, the specific heat below $T^*$ is hardly affected by fields up to 7 Tesla. 
These seem difficult to reconcile with $\mu$SR measurements where the low-temperature muon relaxation rate 
is suppressed sensitively by weak magnetic fields $\sim 10$ mT~\cite{MacLaughlin2008}. 
Thus, it is controversial whether the low-energy thermal excitations are attributable 
to some magnetic excitations, e.g. linear spin waves, or not. 
The origin of the $C_M \propto T^2$ behavior seems still quite elusive. 

On the theoretical side, spin exchanges were estimated 
by a first-principles electronic structure calculation (LDA+U)~\cite{Mazin2007} 
and unrestricted Hartree-Fock (UHF) calculations combined 
with x-ray photo-emission spectroscopy (XPS) ~\cite{Takubo2007, Takubo2009} 
or with Bayesian inference~\cite{Takenaka2014}. 
These are based on the itinerant electron picture. 
They concluded that the third nearest-neighbor coupling is large, 
which naturally leads to the magnetic ordering with ${\mib Q} \sim (1/6, 1/6)$. 
In contrast to these calculations, a more recent ab-initio cluster calculation concludes 
the first nearest-neighbor spin exchange is ferromagnetic and the largest~\cite{Pradines2018}. 
Thus, determination of spin exchanges still remains an unsettled issue. 

Respecting the absence of long-range order of spin dipoles, 
ferro- or antiferro-quadrupole (or spin nematic) ordering has been investigated intensively 
in early studies~\cite{Tsunetsugu2006, Lauchli2006, Bhattacharjee2006}. 
These studies adopt an $S=1$ antiferromagnetic Heisenberg model with the additional biquadratic term, 
based on the localized spin picture. 
When the biquadratic term overcomes the bilinear terms, the spin-nematic or quadrupole ordered state 
becomes the most stable. 
However, these exotic spin-nematic or quadrupole ordered ground states are unlikely to describe 
the real magnetic state of NiGa$_2$S$_4$ at least at the lowest temperature, 
since they yield no magnetic dipole moment and therefore lead to inconsistency 
with the occurrence of an internal magnetic field. 
On the other hand, a later Monte-Carlo study along this line~\cite{Stoudenmire2009} 
gives a possible account for the two humps in specific heat, 
where the low-temperature hump is associated with the $C_3$ bond ordering. 
Therefore it may be possible to regard these exotic quadrupole (or spin nematic) scenarios 
as effectively describe thermally excited states at finite temperatures rather than the ground state. 

In this article, we discuss the magnetic properties of NiGa$_2$S$_4$ 
within conventional mean-field (MF) theory using an itinerant $d$-$p$ model. 
Although our study is similar to the above mentioned early studies 
based on the itinerant picture~\cite{Mazin2007, Takubo2009}, 
it differs from them in the sense that non-collinear spiral states with arbitrary periodicity are assumed 
and the itinerant tight-binding model is constructed by a much less empirical way 
than in the previous Hartree-Fock study. 
Two underlying naive questions driving our study are as follows : 
(i) If the third nearest-neighbor spin exchange and the quadrupole terms 
in the extended Heisenberg Hamiltonian are indeed essential, 
the original itinerant model with long-range electron hoppings should be a more reliable starting point. 
Fortunately, recent first-principles electronic structure calculations enable us 
to construct precise tight-binding models in a less-empirical way, 
based on the so-called maximally localized Wannier functions (MLWF's)~\cite{Mostofi2008, Marzari2012}. 
Starting with such a precise itinerant band model, 
can we explain the nontrivial incommensurate spin configuration ? 
(ii) The observed spin magnetic moment is significantly reduced from $S=1$ with $\delta S \sim - 0.5$, 
which has been ascribed to quantum fluctuations~\cite{Nakatsuji2005, Nakatsuji2010}. 
On the other hand, according to a typical estimation by the standard spin-wave theory, 
the reduction is about $\delta S \sim - 0.26 $ for an antiferromagnetic triangular lattice~\cite{Chernyshev2009}. 
Why is the reduction so large in NiGa$_2$S$_4$? 

Generally, MF theories presuppose occurrence of long-range ordering, 
and therefore may be assumed to be inapplicable to systems where fluctuations are highly active. 
However, still they are useful to investigate what magnetic correlation is the most favorable, 
even if long-range ordering is not completely attained. 
The meaning of our MF calculation is potentially supported by the fact 
that there actually exists a favorable spin configuration characterized by ${\mib Q}$, 
as neutron scattering clearly shows. 
MF theories including the random-phase approximation (RPA), as well as DFT-based calculations, 
only poorly describe finite-temperature properties~\cite{Moriya}. 
Therefore we have to bear in mind that our below discussion holds effectiveness 
only for the lowest-temperature properties, not for finite-temperature properties above 2 K 
where spins keep fluctuating. 

This paper is organized in the following way: 
In \S~\ref{Sc:2.1}, a 17 band $d$-$p$ model for the Ni-$d$ and S-$p$ states is constructed 
from electron band calculation and MLWF's. 
In \S~\ref{Sc:2.2}, the MF theory for spiral states is explained. 
Main calculated results on magnetic properties are presented in \S~\ref{Sc:3}. 
In \S~\ref{Sc:3.1}, it is shown that the most stable spin configuration 
is presented by ${\mib Q} \approx (0.15, 0.15)$, for not so strong $U \approx 2$ eV. 
In \S~\ref{Sc:3.2}, the electronic state under the stable spiral ordering is discussed. 
In \S~\ref{Sc:3.3}, to map onto a classical Heisenberg model, 
nearest-neighbor spin exchange interactions are calculated from the MF result. 
In \S~\ref{Sc:3.4}, the dynamical spin correlation function is calculated within RPA. 
Finally, in \S~\ref{Sc:4}, some brief concluding remarks are given. 

\section{Model Construction and Mean-field Theory} 
\label{Sc:2}

\subsection{Electronic structure}
\label{Sc:2.1}

\begin{figure}
\begin{center}
\includegraphics[width=90mm]{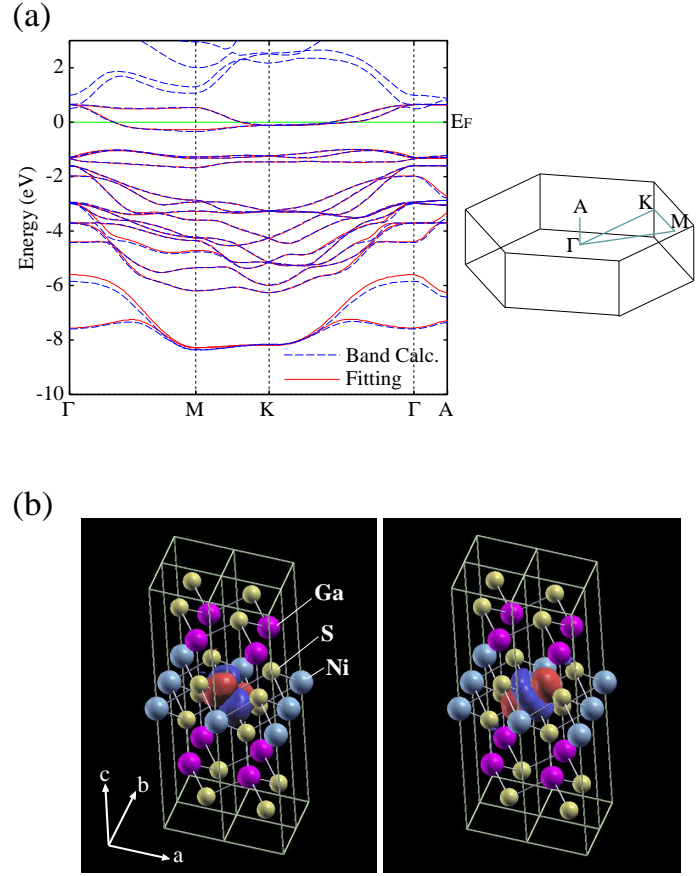}
\end{center}
\caption{(Color online) 
(a) Results of first-principles electronic structure (dashed curves) 
and tight-binding fitting (solid curves) by the MLWF's. The Fermi energy $E_F$ is set to zero. 
The first Brillouin zone and high-symmetry path are depicted in the right-hand side. 
(b) Two degenerate $e_g$-like ($d_{x^2-y^2}$ and $d_{3z^2-r^2}$) MLWF's at a Ni site. }
\label{Fig:1}
\end{figure}

Crystal structure of NiGa$_2$S$_4$ is characterized by layered NiS$_2$ networks which are inter-spaced by GaS ones.  
In a NiS$_2$ layer, nearest eight S atoms surrounding a Ni atom form an approximately regular octahedron. 
Crystal symmetry is represented by a trigonal space group $P\bar{3}m1$, 
where the angle between the $a$ and $b$ axes equals 120$^\circ$, 
and the $c$ axis is perpendicular to the $ab$ plane. 
The lattice parameters are $a=b=0.36249$ nm, and $c=1.19956$ nm~\cite{Lutz1986}. 
Based on the crystal lattice information, we carried out an electronic structure calculation 
for the nonmagnetic state, using WIEN2k~\cite{Blaha}, 
where $32 \times 32 \times 8$ ${\mib k}$-points and 
the Perdew-Burke-Ernzerhof (PBE) exchange-correlation functional are taken. 
The calculated band structure is displayed by the dashed curves in Fig.~\ref{Fig:1}(a). 
Roughly speaking, the five relatively flat bands within the energy region $-1.65$ - $+0.65$eV 
from the Fermi level originate from Ni-$d$ states, 
while the twelve bands within $-8.28$ - $-1.58$ eV originate mainly from S-$p$ states. 
Reflecting the strong two-dimensionality, the energy band dispersion along $\Gamma$-A is weak near the Fermi energy. 

We construct an effective tight-binding model, taking only relevant orbitals into account. 
Since the partial density of states near the Fermi is dominated by the Ni-$d$ and S-$p$ states, 
we take 17 MLWF's (5 Ni-$d$ and 3 $\times$ 4 S-$p$ types) 
to perform tight-binding fitting using the wannier90 code~\cite{Mostofi2008, Marzari2012}. 
To represent the MLWF's, we use the local cartesian coordinate axes defined in the following way: 
The local three-fold symmetry axis [111] is made precisely parallel along the lattice $c$ axis, 
and the local $x$, $y$ and $z$ axes are evenly oriented approximately parallel to the nearest-neighbor Ni-S bonds. 
Since the S atoms form a not completely regular but slightly distorted octahedron centered by a Ni atom, 
the $x$, $y$ and $z$ axes do not precisely point to the center of the surrounding S atoms. 
However, this choice is convenient, allowing us to discuss the electronic properties 
through the conventional view based on a local octahedral ligand field. 
By this choice, while $t_{2g}$ states are almost completely filled, 
$e_g$ states are partly filled with electrons, as seen later in Table~\ref{Tab:1}. 
As a result of band fitting, we have precisely three-fold degenerate $t_{2g}$-type ($xy$, $yz$, $xz$) MLWF's 
and two-fold degenerate $e_g$-type ($x^2-y^2$ and $3z^2-r^2$) MLWF's at each Ni site. 
The two $e_g$-type MLWF's are depicted in Fig.~\ref{Fig:1}(b). 
The tight-binding bands are shown by the solid curves in Fig.~\ref{Fig:1}(a). 
The obtained tight-binding model has the form of 
\begin{eqnarray}
H_0 & = & \sum_{i\ell}^{\rm @Ni} \sum_{i'\ell'}^{\rm @Ni} \sum_\sigma 
t_{i \ell, i'\ell'} d_{i \ell \sigma}^{\dag} d_{i' \ell' \sigma} 
+ \sum_{jm}^{\rm @S} \sum_{j'm'}^{\rm @S} \sum_\sigma 
t_{jm,j'm'} p_{j m \sigma}^{\dag} p_{j' m' \sigma} \nonumber\\
&& + \sum_{i \ell}^{\rm @Ni} \sum_{j m}^{\rm @S} \sum_\sigma 
(t_{i \ell, j m} d_{i \ell \sigma}^{\dag} p_{j m \sigma} + h.c.), 
\end{eqnarray}
where $d_{i \ell \sigma}^{\dag}$ and $d_{i \ell \sigma}$ 
with $\ell = xy, yz, xz, x^2-y^2, 3z^2-r^2$ 
($p_{j m \sigma}^{\dag}$ and $p_{j m \sigma}$ with $m=x,y,z$) 
are the creation and annihilation operators for the Ni-$d_\ell$ (S-$p_m$) electrons 
at Ni site $i$ (S site $j$) with spin $\sigma$. 
One-particle energies of Ni-$d_\ell$ and S-$p_m$ states are 
$\varepsilon_\ell \equiv t_{i\ell,i\ell}$ and $\varepsilon_m \equiv t_{jm,jm}$. 
For the Ni-$d$ orbitals, $\varepsilon_{t_{2g}} = -1.58$ eV 
and $\varepsilon_{e_g} = -1.32$ eV with respect to the Fermi level. 
Some of calculated transfer integrals along nearest-neighbor $\sigma$-bonding are, 
e.g., $t_{3z^2-r^2, z} = 1.00$ eV between the Ni-$d_{3z^2-r^2}$ and S-$p_z$ orbitals, 
and $t_{x^2-y^2, x(y)} = 0.87$ eV between the Ni-$d_{x^2-y^2}$ and S-$p_{x(y)}$ orbitals. 

\subsection{Spiral mean-field theory}
\label{Sc:2.2}

The $d$-$p$ Hamiltonian for the Ni-$d$ and S-$p$ states has the following form: 
\begin{equation}
H = H_0 + H'
\end{equation}
where the non-interacting part $H_0$ is determined already in \S~\ref{Sc:2.1}. 
$H'$ is the Coulomb interaction at Ni sites: 
\begin{eqnarray}
H' & = & \sum_i^{\rm @Ni} \biggl[ \frac{U}{2} \sum_{\ell} \sum_{\sigma \neq \sigma'} 
d_{i \ell \sigma}^{\dag} d_{i \ell \sigma'}^{\dag} d_{i \ell \sigma'} d_{i \ell \sigma} 
+ \frac{U'}{2} \sum_{\ell \neq \ell'} \sum_{\sigma \sigma'} 
d_{i \ell \sigma}^{\dag} d_{i \ell' \sigma'}^{\dag} d_{i \ell' \sigma'} d_{i \ell \sigma} \nonumber\\
&& + \frac{J}{2} \sum_{\ell \neq \ell'} \sum_{\sigma \sigma'}
d_{i \ell \sigma}^{\dag} d_{i \ell' \sigma'}^{\dag} d_{i \ell \sigma'} d_{i \ell' \sigma} 
+ \frac{J}{2} \sum_{\ell \neq \ell'} \sum_{\sigma \neq \sigma'}
d_{i \ell \sigma}^{\dag} d_{i \ell \sigma'}^{\dag} d_{i \ell' \sigma'} d_{i \ell' \sigma} \biggr] , 
\end{eqnarray}
We apply the MF approximation to $H'$: 
\begin{eqnarray}
H'_{MF} &=& \sum_{i, \ell}^{{\rm @Ni}} \biggl[ \frac{U}{2} \langle n_{i\ell} \rangle 
+ \biggl (U' - \frac{J}{2}  \biggr) \sum_{\ell' (\neq \ell)} 
\langle n_{i\ell'} \rangle \biggr] n_{i\ell} \nonumber\\ 
&& - \sum_{i,\ell}^{{\rm @Ni}} \biggl[ \frac{U}{2} \langle \mib{m}_{i\ell} \rangle 
+ \frac{J}{2} \sum_{\ell'(\neq \ell)} \langle \mib{m}_{i\ell'} \rangle \biggr] 
\cdot \mib{m}_{i\ell} - \sum_{i, \ell}^{{\rm @Ni}} \frac{U}{4} 
\biggl[ \langle n_{i\ell} \rangle^2 - | \langle \mib{m}_{i\ell} \rangle |^2 \biggr] \nonumber \\
&& - \sum_{i, \ell \neq \ell'}^{{\rm @Ni}} 
\frac{U'}{2} \langle n_{i\ell} \rangle \langle n_{i\ell'} \rangle 
+ \sum_{i, \ell \neq \ell'}^{\rm @Ni} \frac{J}{4} 
\biggl[ \langle n_{i\ell} \rangle \langle n_{i\ell'} \rangle 
+ \langle \mib{m}_{i\ell} \rangle \cdot \langle \mib{m}_{i\ell'} \rangle \biggr], 
\end{eqnarray}
where $\langle n_{i\ell} \rangle$ and $\langle \mib{m}_{i\ell} \rangle$ are the mean-fields of 
\begin{eqnarray}
n_{i\ell} &=& \sum_{\sigma} d_{i\ell\sigma}^{\dag} d_{i\ell\sigma}, \\
\mib{m}_{i\ell} &=& \sum_{\sigma\sigma'} d_{i\ell\sigma}^{\dag} 
\mib{\sigma}_{\sigma\sigma'} d_{i\ell\sigma'}, 
\end{eqnarray}
with $\mib{\sigma}$ the Pauli matrix vector. 
Here we assume the spiral-ordering spins are confined parallel to the $ab$ plane 
and furthermore expressed by a single pitch vector ${\mib Q}$ : 
\begin{equation}
\langle \mib{m}_{i\ell} \rangle = \bigl( \langle m_{i\ell}^x \rangle, \langle m_{i\ell}^y \rangle, 0 \bigr) 
= \bigl| \langle \mib{m}_{{\mib Q} \ell} \rangle \bigr| \bigl( \cos[{\mib Q} \cdot {\mib r}_i + \phi], 
\sin[{\mib Q} \cdot {\mib r}_i + \phi], 0 \bigr). 
\end{equation}
It is more convenient to use 
\begin{equation}
\langle m_{{\mib Q} \ell}^+ \rangle \equiv \langle m_{i\ell}^x \rangle + i \langle m_{i\ell}^y \rangle
= | \langle m_{{\mib Q}\ell}^+ \rangle | e^{i({\mib Q} \cdot {\mib r}_i + \phi)}, 
\end{equation}
where 
\begin{equation}
m_{{\mib Q}\ell}^+ 
= \frac{1}{N} \sum_{\mib k} \sum_{\sigma \sigma'} 
d_{{\mib k} \ell \sigma}^{\dag} [\sigma_+]_{\sigma \sigma'} d_{{\mib k}+{\mib Q} \ell \sigma'} 
= \frac{2}{N} \sum_{\mib k} \sum_{\sigma \sigma'} 
d_{{\mib k} \ell \uparrow}^{\dag} d_{{\mib k}+{\mib Q} \ell \downarrow}, 
\end{equation}
with $\sigma_+ = \sigma_x + i \sigma_y$. 
$d_{{\mib k} \ell \sigma}^{\dag}$ and $d_{{\mib k} \ell \sigma}$ 
are the Fourier transforms of $d_{i \ell \sigma}^{\dag}$ and $d_{i \ell \sigma}$, respectively. 
For charge density, we assume the spatially uniform case: 
\begin{equation}
\langle n_{i\ell} \rangle = \langle n_\ell \rangle. 
\end{equation}
Under these assumptions, $H'_{MF}$ is expressed in momentum representation as 
\begin{eqnarray}
H'_{MF} &=& \sum_{{\mib k} \ell \sigma} \biggl[ \frac{U}{2} \langle n_{\ell} \rangle 
+ \biggl (U' - \frac{J}{2}  \biggr) \sum_{\ell' (\neq \ell)} \langle n_{\ell'} \rangle \biggr] 
d_{{\mib k} \ell \sigma}^{\dag} d_{{\mib k} \ell \sigma} \nonumber\\ 
&& - \sum_{{\mib k} \ell} \biggl\{ \biggl[ \frac{U}{2} \langle m_{{\mib Q}\ell}^+ \rangle 
+ \frac{J}{2} \sum_{\ell'(\neq \ell)} \langle m_{{\mib Q}\ell'}^+ \rangle \biggr]^* 
d_{{\mib k} \ell \uparrow}^{\dag} d_{{\mib k}+{\mib Q} \ell \downarrow} + h.c. \biggr\} \nonumber\\
&& - \frac{NU}{4}  \sum_{\ell} \biggl[ \langle n_{\ell} \rangle^2 
- | \langle m_{{\mib Q}\ell}^+ \rangle |^2 \biggr] - \frac{NU'}{2} \sum_{\ell \neq \ell'} 
\langle n_\ell \rangle \langle n_\ell' \rangle \nonumber\\
&& + \frac{NJ}{4}  \sum_{\ell \neq \ell'}\biggl[ \langle n_\ell \rangle \langle n_{\ell'} \rangle 
+ {\rm Re} \{ \langle m_{{\mib Q}\ell}^+ \rangle \langle m_{{\mib Q}\ell'}^+ \rangle^* \} \biggr]. 
\end{eqnarray}
The total MF Hamiltonian $H_{MF} \equiv H_0 + H'_{MF} $ can be easily diagonalized in momentum representation, 
by introducing new fermionic operators $c_{{\mib k}a}$ and diagonalization matrix $u_{\ell\sigma, a}({\mib k})$: 
\begin{equation}
[ d_{{\mib k} \ell\uparrow}, d_{{\mib k}+{\mib Q} \ell\downarrow}, 
p_{{\mib k} m \uparrow}, p_{{\mib k}+{\mib Q}   m \downarrow} ] 
= \sum_a [ u_{\ell\uparrow, a}({\mib k}), u_{\ell\downarrow, a}({\mib k}), 
u_{m \uparrow, a}({\mib k}), u_{m \downarrow, a}({\mib k}) ] c_{{\mib k}a}. 
\end{equation}
We have 34 energy bands ($E_a(\mib{k})$, $1 \leq a \leq 34$) 
in the spiral magnetic ground state. 

Within the MF theory, we should consider that the one-particle energy 
$\varepsilon_{\ell}$ already includes the following energy shift from the bare one, 
due to the electron-electron Coulomb interaction at Ni site. 
Therefore, before determining the magnetic ground state, 
we need to evaluate the bare one-particle energy by 
$\varepsilon_{\ell}^{(0)} \equiv \varepsilon_{\ell} - \Delta \varepsilon_\ell$, 
where $\Delta \varepsilon_\ell$ is calculated from the expectation values 
of particle numbers $\langle n_{i\ell} \rangle$'s in the nonmagnetic state by 
\begin{equation}
\Delta \varepsilon_\ell \equiv \biggl[ \frac{U}{2} \langle n_{\ell} \rangle 
+ \biggl (U' - \frac{J}{2}  \biggr) \sum_{\ell' (\neq \ell)} \langle n_{\ell'} 
\rangle \biggr]_{\rm nonmag.}. 
\end{equation}
Maintaining values of $\varepsilon_{\ell}^{(0)}$, we determine the mean-fields 
$\langle n_\ell \rangle$ and $\langle m_{{\mib Q}\ell}^+ \rangle$, 
by solving the self-consistency equations: 
\begin{eqnarray}
\langle n_{\ell} \rangle &=& \frac{1}{N} \sum_{{\mib k} a \sigma} 
u_{\ell\sigma, a}^*({\mib k}) u_{\ell\sigma, a}({\mib k}) f(E_a({\mib k})), \\
\langle m_{{\mib Q}\ell}^+ \rangle &=& \frac{2}{N} \sum_{{\mib k} a} 
u_{\ell\uparrow, a}^*({\mib k}) u_{\ell\downarrow, a}({\mib k}) f(E_a({\mib k})), 
\end{eqnarray}
where $f(z) = 1/[e^{(z-\mu)/T} + 1]$ is the Fermi distribution function 
(Chemical potential $\mu$ is always adjusted to maintain the total electron number 
and $T = 0.01 {\rm K} \approx 8.62 \times 10^{-7}$ eV for below calculations). 
Numerical integrations in ${\mib k}$ are carried out using oblique ${\mib k}$-meshes, 
where the parallel-piped reciprocal unit cell is divided into $84 \times 84 \times 8$ 
${\mib k}$-meshes for most below results (Some results are by $168 \times 168 \times 8$ or 
$168 \times 168 \times 16$ ${\mib k}$-meshes). 

Energy per unit cell is calculated by 
\begin{eqnarray}
\frac{\langle H_{MF} \rangle}{N} &=& \frac{1}{N} \sum_{{\mib k}, a} E_a({\mib k}) f(E_a({\mib k}))
- \frac{U}{4}  \sum_{\ell} \biggl[ \langle n_{\ell} \rangle^2 
- | \langle m_{{\mib Q}\ell}^+ \rangle |^2 \biggr] - \frac{U'}{2} \sum_{\ell \neq \ell'} 
\langle n_\ell \rangle \langle n_\ell' \rangle \nonumber\\
&& + \frac{J}{4}  \sum_{\ell \neq \ell'}\biggl[ \langle n_\ell \rangle \langle n_{\ell'} \rangle 
+ {\rm Re} \{ \langle m_{{\mib Q}\ell}^+ \rangle \langle m_{{\mib Q}\ell'}^+ \rangle^* \} \biggr]. 
\end{eqnarray}
The stabilization energy due to magnetic ordering is calculated by subtracting 
$\langle H_{MF} \rangle / N$ of the nonmagnetic state from that of the magnetic state. 

\section{Results on Magnetic Properties}
\label{Sc:3}

\subsection{Search for the most stable spin configuration}
\label{Sc:3.1} 

\begin{figure}
\begin{center}
\includegraphics[width=120mm]{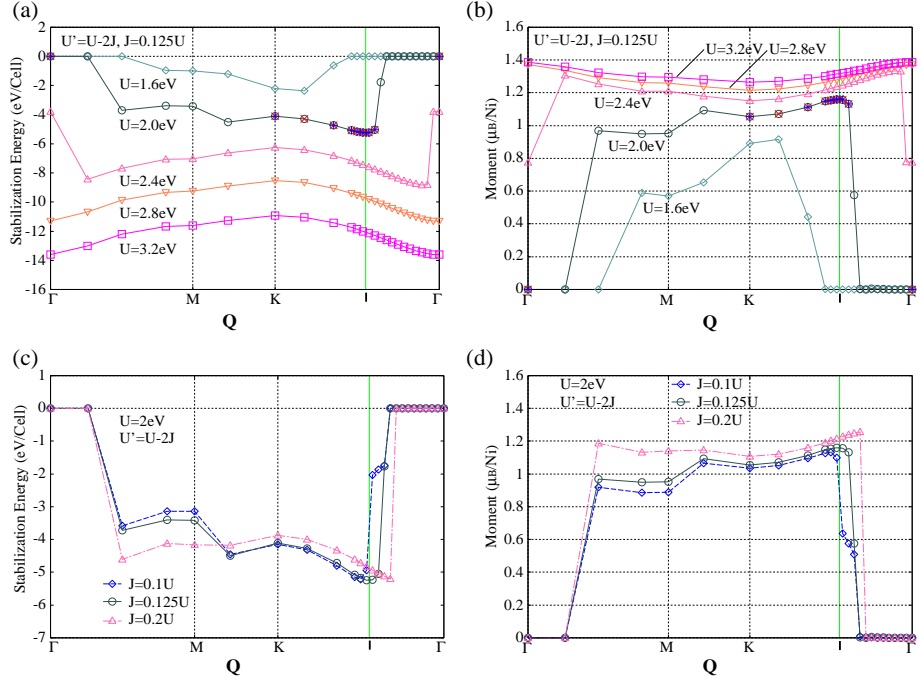}
\end{center}
\caption{(Color online) 
(a) Calculated stabilization energies (per unit cell) along the symmetry line path $\Gamma$-$M$-$K$-$\Gamma$ 
for $U=1.6$ eV, 2.0 eV, 2.8 eV, and 3.2 eV with $U'=U-2J$ and $J=0.125U$. 
(b) Calculated total spin moments at each Ni site for the same Coulomb interaction parameters and ${\mib Q}$'s as in (a). 
(c) Calculated stabilization energies along the same symmetry line path 
for $J=0.1U$, $0.125U$ and $0.2U$ with $U'=U-2J$ and $U=2.0$ eV. 
(d) Calculated total spin moments at Ni site for the same Coulomb interaction parameters and ${\mib Q}$'s as in (c). 
Note that $\Gamma$, $M$ and $K$ correspond to the uniform ferromagnetic, the ``stripe" antiferromagnetic 
and the 120$^\circ$ orderings, respectively. 
Empty symbols are the results calculated by $84 \times 84 \times 8$ ${\mib k}$-meshes. 
On the $U=2.0$ eV curve in (a) and (b), $\times$ ($+$) at $K$, $\Gamma$ and several points 
between them indicates the results for $168 \times 168 \times 8$ ($168 \times 168 \times 16$) ${\mib k}$-meshes. 
The vertical solid line between $K$ and $\Gamma$ indicates ${\mib Q}=(0.15,0.15)$.}
\label{Fig:2}
\end{figure}

To find out a spiral ordering vector giving the ground state, 
we calculate MF self-consistent solutions along the symmetry path $\Gamma$-$M$-$K$-$\Gamma$ 
(Hereafter we restrict ourselves to the cases of two-dimensional ${\mib Q}$, 
i.e., $Q_c = 0$, and show only in-plane components ${\mib Q} = (Q_a, Q_b)$ explicitly). 
Results of calculated stabilization energies and the spin moments for various ${\mib Q}$'s 
are displayed in Fig.~\ref{Fig:2}. 
At weak $U = 1.6$ eV, we find the magnetic ground state near ${\mib Q}=K=(1/3,1/3)$ between $\Gamma$ and $K$, 
while we find no magnetic solutions around ${\mib Q}=\Gamma=(0,0)$. 
Here note that $\Gamma$ and $K$ correspond to the uniform ferromagnetic ordering 
and the 120$^\circ$ ordering, respectively. 
As $U$ is increased, the magnetic correlation tends toward the uniform ferromagnetic 
rather than toward the 120$^\circ$ ordering, as shown in Fig.~\ref{Fig:2}(a). 
Before reaching the uniform ferromagnetic ordering, for $U \approx 2.0$ eV, 
we have the spiral ordering with ${\mib Q} \approx (0.15, 0.15)$ for the most stable state, 
which is consistent with neutron scattering. 
In this spiral ground state, the spin magnetic moment equals about 1.16 $\mu_B$, 
which corresponds to $ |\langle S \rangle| \approx 0.58$, 
being not away from the experimental value $\sim 0.51$. 
If $U$ is increased to be larger than 2.8 eV, ${\mib Q}$ giving the most stable state becomes 
fixed at $\Gamma$ (uniform ferromagnetic state). 
Thus, the magnetic correlation is predominantly ferromagnetic rather than antiferromagnetic. 

To see effects of the Hund's rule coupling, we present the results 
for some different values of $J$ in Fig.~\ref{Fig:2}(c) and (d). 
The results suggest that $J$ does not affect drastically the stabilization energy 
and the magnitude of the magnetic moment, as far as we restrict $J/U$ to a realistic value 0.1 - 0.2. 

To understand these magnetic properties mentioned in this section, 
we shall take a close look into the electronic states of the spiral magnetic state 
as well as the nonmagnetic state, in the next section. 

\subsection{Electron states in the magnetic ground state}
\label{Sc:3.2}

Generally, one of crucial factors determining magnetic correlations in transition-metal compounds 
is electron fillings of transition-metal $d$ states~\cite{Moriya, Alexander1964, Moriya1965}. 
The electron configurations for the nonmagnetic and the spiral magnetic (${\mib Q}=(0.15, 0.15)$) states 
are shown in Table.~\ref{Tab:1}. 
\begin{table}
\caption{Ni-$d$ electron configuration in the nonmagnetic and magnetic states. 
$n$ and $m$ are the total Ni-$d$ electron number and spin moment (in units of $\mu_B$) 
per Ni site. }
\begin{tabular}{ll ccccc rr}
\hline
              &   & $xy$ & $yz$ & $xz$ & $x^{2}-y^{2}$ & $3z^{2}-r^{2}$ & $n$ & $m$ \\
\hline
Nonmagnetic       & $\uparrow$   & 0.991 & 0.991 & 0.991 & 0.745 & 0.745 & 8.93 & 0.00 \\
                  & $\downarrow$ & 0.991 & 0.991 & 0.991 & 0.745 & 0.745 &      & \\
\hline
Magnetic          & $\uparrow$   & 0.998 & 0.996 & 0.998 & {\bf 0.981} & {\bf 0.977} & {\bf 8.74} & {\bf 1.16} \\
(${\mib Q}=(0.15,0.15)$) & $\downarrow$ & 0.992 & 0.995 & 0.992 & {\bf 0.391} & {\bf 0.416} &            & \\
\hline
\end{tabular}
\label{Tab:1}
\end{table}
In the nonmagnetic state, the Ni-$t_{2g}$ states are almost completely filled with electrons, 
while the $e_g$ states are filled by 75 \%. 
This non-integral electron filling in the $e_g$ state reflects the covalent bonding 
between Ni-$e_g$ and ligand S-$p$ orbitals. 
We consider this quarter filling (in hole representation) of the $e_g$ states 
is a reason for the predominant ferromagnetic correlation. 
The microscopic mechanism of this ferromagnetic correlation is basically 
the same as explained in Refs.~\cite{Alexander1964, Moriya1965} 
and more clearly in \S 6.6 of Ref.~\cite{Moriya}. 
Thus the present weak-coupling analysis provides another view quite different from the following view 
from the localized ionic picture, 
``Each Ni$^{2+}(d^8)$ ion should have two holes, whose total spin moment should be 2 $\mu_B$ ($S=1$). 
Then, the two-fold degenerate $e_g$ states are evenly filled just by half, 
and therefore the spin correlation between the nearest-neighboring Ni sites should be antiferromagnetic." 
Note that the valence of Ni is closer to $d^9$ rather than to $d^8$. 
In fact, a model calculation to analyze Ni2$p_{3/2}$ XPS spectra indicated 
the ground state has the $d^9L$ character ($L$ is a S $3p$ hole), 
although much larger $U$ (5.0 eV) was used there~\cite{Takubo2007}.

For the spiral magnetic state, the Ni-$t_{2g}$ states are almost completely occupied with electrons again. 
This means the $t_{2g}$ states do not play any important role in low-energy electronic properties. 
The total number of $e_g$ electrons are about 2.77 (69 \% filled), still significantly deviating from half filling. 
Thus, for this electron filling, even though the Ni-$d$ states were completely polarized in spin, 
the magnitude of the spin moment could reach at most 1.23 $\mu_B$ ($|\langle S \rangle|=0.62$), 
due to the covalency between Ni-$e_g$ and S-$p$ states. 
We consider this can be an origin of the significant reduction of Ni spins, 
alternative or additional to fluctuation effects. 

In Table.~\ref{Tab:1}, we can see a small difference of electron occupation number 
between the $x^{2}-y^{2}$ and $3z^{2}-r^{2}$ orbitals in the spiral magnetic state. 
This is because the spiral ordering state makes the electronic structure lose the $C_3$ rotational symmetry. 
We confirmed that this filling difference is so small that substantial results are not affected 
by the initial choice of the local $x,y$ and $z$ axes for defining the local orbitals. 
Therefore we can conclude that clear orbital ordering is unlikely to accompany 
the spiral magnetic ordering in NiGa$_2$S$_4$. 

\begin{figure}
\begin{center}
\includegraphics*[width=90mm]{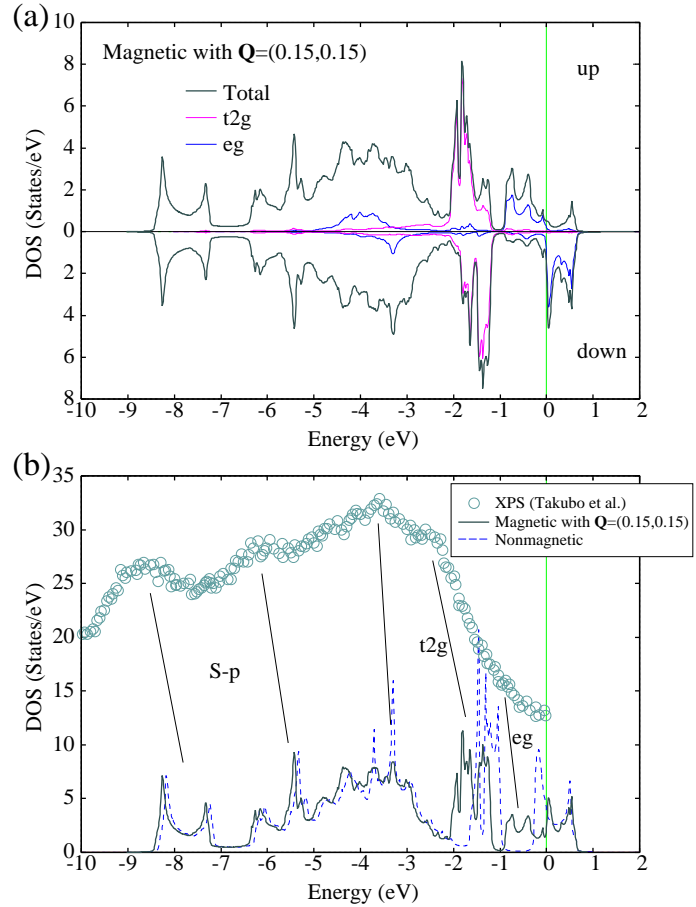}
\end{center}
\caption{(Color online) 
(a) Calculated density of states (DOS) for up- and down-spin states 
for the magnetic ordered state with ${\mib Q}=(0.15, 0.15)$. 
Thin curves represent the partial DOS of Ni-$d$ $t_{2g}$ and $e_g$ states. 
(b) The total DOS summed in spin is compared with the XPS data read from ref.~\cite{Takubo2007, Takubo2009}. }
\label{Fig:3}
\end{figure}

Calculated density of states (DOS) of the magnetic state is presented in Fig.~\ref{Fig:3}(a) and (b). 
Overall qualitative agreement between the calculated DOS and XPS data suggests 
that the original band structure calculation and the model derived from it 
could capture the real electronic structure. 
Calculated energy levels of the $t_{2g}$ states are somewhat near the Fermi level, 
compared with the experiment, maybe due to underestimation of $10Dq \equiv \varepsilon_{e_g} - \varepsilon_{t_{2g}}$. 
As far as we discuss electronic properties at low energies, this underestimation will not give rise to crucial problems, 
since the $t_{2g}$ states are fully occupied throughout. 
This full occupancy of the $t_{2g}$ states is responsible for the ineffectiveness of the Hund's coupling: 
Since the $t_{2g}$ states are not polarized in spin at all, the Hund's coupling works only 
between the two $e_g$ states to make their spins parallel. 

One may consider that a small but finite DOS remaining at the Fermi level contradicts 
resistivity and spectroscopic experiments which suggest insulating behaviors 
and the existence of energy gap of 0.2 - 0.3 eV~\cite{Takubo2009,Tomita2009}. 
As generally admitted, most itinerant approaches underestimate local electron correlations, 
and this underestimation makes insulating gap tend to close in calculations. 
Therefore the absence of energy gap may be a possible defect of the theoretical approach. 
Nevertheless, we shall raise below some reasons why we still consider 
it worthwhile to maintain the itinerant description: 
First, calculated electron numbers of partly filled bands in the spiral magnetic state 
are 0.986 and 0.014 for the 32nd and 33rd bands, respectively. 
This means the Fermi surface volume should be negligibly small, only about 1-2\% 
of the Brillouin zone and the system can behave virtually like an insulator 
or at most like low-carrier semi-metal. 
Furthermore, in reality, the present spiral magnetic ordering allows six kinds 
of magnetic domains, whose boundaries could disturb metallic transport, 
making the system tend toward insulator, possibly accompanied by some gapful behavior. 
Photo-emission spectroscopy may have difficulty in bulk sensitivity and resolution 
for detecting an extremely small DOS near $E_F$. 
Thus we should not conclude simply that the calculated electronic state 
is so much unlike the real electronic state that the effectiveness 
of the itinerant description is excluded. 

\subsection{Mapping onto classical Heisenberg model}
\label{Sc:3.3}

In general, Heisenberg spin models are valid only for such completely localized electron systems 
as Mott insulators with $U \gg t$. 
Therefore, it is unclear how legitimate mapping onto Heisenberg spin description 
is in the present weak coupling case. 
However, plausible techniques have been developed to estimate the effective Heisenberg spin exchanges 
even for itinerant magnets~\cite{Lacour-Gayet1974, Oguchi1983, Liechtenstein1987}. 
Now we assume the low-energy magnetic properties of spiral-ordered spins 
can be described by the classical Heisenberg model: 
\begin{equation}
H_S = - \sum_{(n,i)}^{\rm @Ni} \sum_{\mu,\nu = x,y,z} J_{\mu\nu}(n,i) e_\mu(n) e_\nu(i), 
\end{equation}
where $e_\mu(i)$ is the $\mu$ component of the unit vector pointing along the spin moment at Ni site $i$. 
Then the spin exchange interaction $J_{\mu\nu}(n,i)$ is calculated by~\cite{Liechtenstein1987} 
\begin{equation}
J_{\mu\nu}(n,i)= \sum_{\ell\ell'} \frac{1}{4\pi} \int_{-\infty}^{\infty} dz\, f(z) 
\Delta_{n\ell} \Delta_{i\ell'} {\rm Im} {\rm Tr} 
[\sigma_\mu G_{n\ell, i\ell'}(z) \sigma_\nu G_{i\ell', n\ell}(z) ], 
\label{Eq:Jni}
\end{equation}
where $f(z)$ is the Fermi function, $G_{n\ell, i\ell'}(z)$ is the real-space Green's function 
($2 \times 2$ matrix in spin space) in the magnetic state, trace summation is for spin indices, 
and $\Delta_{i\ell}$ is the magnetic exchange splitting of $d_\ell$ state at Ni site $i$: 
\begin{equation}
\Delta_{i\ell} \equiv \biggl| U \langle \mib{m}_{i\ell} \rangle 
+ J \sum_{\ell'(\neq \ell)} \langle \mib{m}_{i\ell'} \rangle \biggr|. 
\end{equation}
If we apply this formula within the MF theory to the half-filled case of single-band Hubbard models, 
we can verify straightforwardly that the nearest-neighbor spin exchange correctly 
tends to $-t^2/U$ asymptotically in the strong-coupling limit of $U \rightarrow \infty$~\cite{Lacour-Gayet1974}. 

\begin{figure}
\begin{center}
\includegraphics[width=60mm]{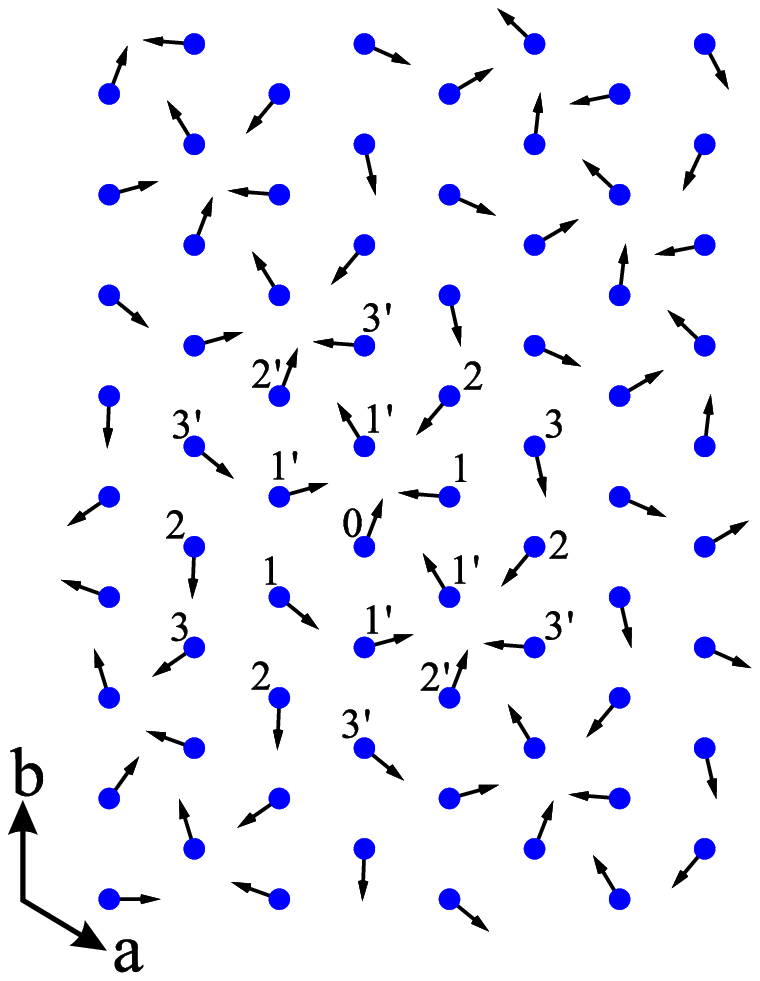}
\begin{tabular}{lccc}
\hline
 $n$  & 1   & 2    & 3    \\
\hline
$J_n$    [meV] & 7.7 & -4.0 & -3.3 \\
$J_{n'}$ [meV] & 4.8 &  2.9 & -2.7 \\
\hline
LDA+U~\cite{Mazin2007}   &  -8.4 & -0.3 &  -4.1 \\
UHF~\cite{Takubo2009}    & -12.0 & -0.3 & -17.9 \\
UHF+B~\cite{Takenaka2014} & -0.08 & -0.23 & -10.57 \\
DDCI2~\cite{Pradines2018} &  1.60 & 0.04 & -0.53 \\
\hline
NS~\cite{Stock2010}      & 1.03$^a$ & & -2.8(6) \\
ESR~\cite{Yamaguchi2010} & 0.39$^b$ & & -1.96$^c$ \\
\hline
\multicolumn{4}{r}{$^a$-0.35(9)$J_3$, $^b$4.56 [K], $^c$-22.8 [K].}
\end{tabular}
\end{center}
\caption{(Color online) 
Calculated spiral spin configuration with ${\mib Q}=(0.15, 0.15)$ is illustrated, 
where filled circles and arrows represent Ni atoms and spin moments, respectively. 
Numbers indicate how far the numbered sites are from the site numbered 0, i.e., 
the 1 and 1' sites are the first nearest, the 2 and 2' sites are the second nearest, 
and the 3 and 3' are the third nearest, from the site 0. 
In the bottom table, $J_{n(n')}$ are the spin exchanges calculated in the present study, in units of meV. 
Positive (negative) value corresponds to ferromagnetic (antiferromagnetic) coupling. 
For comparison, also the values of $J_n$ [meV] estimated in previous studies are listed. 
LDA+U: LDA+U calculation~\cite{Mazin2007}, 
UHF: unrestricted Hartree-Fock calculation~\cite{Takubo2009}, 
UHF+B: Bayesian inference from the UHF results~\cite{Takenaka2014}, 
DDCI2: ab-initio cluster calculation~\cite{Pradines2018}, 
NS: neutron scattering~\cite{Stock2010}, 
ESR: electron spin resonance~\cite{Yamaguchi2010}. } 
\label{Fig:4}
\end{figure}

Calculated results of in-plane components of spin exchanges 
($J_n \equiv J_{ab}(n,0)$ averaged over in-plane spin directions) 
are displayed in the bottom table in Fig.~\ref{Fig:4}, 
where we have to classify neighboring sites into more kinds than in the nonmagnetic state, 
since the incommensurate spiral ordering breaks the original $C_3$ rotational symmetry 
of the electronic structure around the $c$-axis. 
We give a general account for the necessity of such classification 
in the magnetic ordered states in the Appendix. 
In Fig.~\ref{Fig:4}, we have used a prime for the classification, 
by which the $0$-$n'$ directions deviate from the $\pm {\mib Q}$ direction 
more than the $0$-$n$ directions deviate. 
By this classification, we have $|J_n|>|J_{n'}|$, as the numerical results show. 
This is because virtual exchange (hopping) processes between the 0 and $n'$ sites 
are relatively suppressed, compared with those along the $0$-$n$ directions. 

Actually, numerical estimation of $J_n$'s is difficult in precision, 
and we do not exclude the possibility that our numerical values of $J_{n(n')}$ 
can include deviation (at most $\pm 1$ meV) from the true value. 
However, within this precision, we can still stress that the nearest-neighbor couplings 
are ferromagnetic ($J_{1(1')}>0$), while the third nearest-neighbor couplings are antiferromagnetic ($J_{3(3')}<0$). 
This agrees qualitatively with the estimations from neutron scattering~\cite{Stock2010} and electron spin resonance (ESR) 
measurements~\cite{Yamaguchi2010}, and with a recent ab-initio cluster calculation~\cite{Pradines2018}. 
However, the present calculation shows that the first nearest-neighbor coupling should be the largest, 
in contrast to the experimental evaluations. 
We calculated also longer-range exchanges up to the ninth nearest-neighbors, 
which are all smaller than at most 1.2 meV. 

\subsection{Dynamical spin correlations}
\label{Sc:3.4}

\begin{figure}
\begin{center}
\includegraphics[width=90mm]{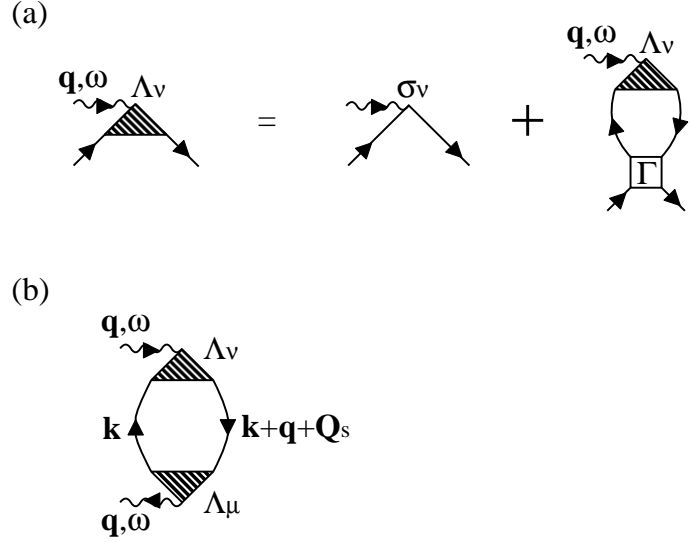}
\end{center}
\caption{
(a) Diagrammatic representation of the random-phase approximation (RPA) 
for the scattering function $\Lambda_{\nu,\zeta_3\zeta_4}({\mib q}, \omega; {\mib Q}_s)$. 
$\Gamma$ and oriented solid lines represent the on-site Coulomb interaction (antisymmetrized) 
and the Green's function (in the magnetic state), respectively. 
(b) Dynamical spin correlation function $S_{\mu\nu}({\mib q}, \omega)$. }
\label{Fig:5}
\end{figure}

For the spiral MF ground state, we calculate the spin excitation spectra, 
which should be compared with neutron scattering. 
In the spiral magnetic state, we need to take Umklapp processes into consideration, 
which yield a momentum shift by ${\mib Q}_s \equiv s{\mib Q} = \{{\mib 0}, \pm {\mib Q} \}$ ($s=0, \pm$) 
in the intermediate and final states. 
We calculate the scattering vertex within the RPA, 
diagrammatically as shown in Fig.~\ref{Fig:5}(a) and analytically as presented in the following : 
\begin{eqnarray}
\Lambda_{\nu, \zeta_3\zeta_4}({\mib q}, \omega; {\mib Q}_s) 
&=& \delta_{\ell_3\ell_4} [\sigma_\nu]_{\sigma_3\sigma_4} 
\delta_{{\mib Q}_s, {\mib Q}_{\sigma_4\sigma_3}} \nonumber\\
&& + \sum_{\zeta_1\zeta_2}{}' \sum_{\xi_3\xi_4} \Gamma_{\zeta_1\zeta_3,\zeta_2\zeta_4} 
\chi_{\zeta_1\zeta_2,\xi_3\xi_4}({\mib q}, \omega; {\mib Q}_s) 
\Lambda_{\nu, \xi_3\xi_4}({\mib q}, \omega; {\mib Q}_s), 
\end{eqnarray}
where we have used compact notation $\zeta_i \equiv (\ell_i, \sigma_i)$, 
${\mib Q}_{\sigma\sigma'} \equiv 
({\mib Q}_{\uparrow\uparrow}, {\mib Q}_{\uparrow\downarrow}, 
{\mib Q}_{\downarrow\uparrow}, {\mib Q}_{\downarrow\downarrow}) \equiv 
({\mib 0}, -{\mib Q}, {\mib Q}, {\mib 0}) $, $\Gamma_{\zeta_1\zeta_3,\zeta_2\zeta_4} $ 
is the antisymmetrized Coulomb interaction vertex taking a value of $\pm U, \pm U'$ or $\pm J$, 
the summation in $\sigma_1$ and $\sigma_2$ with a prime should be taken under the condition 
${\mib Q}_{\sigma_1\sigma_2}={\mib Q}_{\sigma_4\sigma_3}$, 
and $\chi_{\zeta_1\zeta_2,\zeta_3\zeta_4}({\mib q}, \omega; {\mib Q}_s)$ is 
the irreducible susceptibility (in the magnetic state) calculated by 
\begin{eqnarray}
\chi_{\zeta_1\zeta_2,\zeta_3\zeta_4}({\mib q}, \omega; {\mib Q}_s) &=& 
\frac{1}{N} \sum_{\mib k} \sum_{aa'} u_{\zeta_4,a}({\mib k}) u_{\zeta_1,a}^*({\mib k}) 
u_{\zeta_2,a'}({\mib k}+{\mib q}+{\mib Q}_s) u_{\zeta_3,a'}^*({\mib k}+{\mib q}+{\mib Q}_s) \nonumber\\
&& \times \chi_{aa'}({\mib k}; {\mib q},\omega; {\mib Q}_s), \\ 
\chi_{aa'}({\mib k}; {\mib q},\omega; {\mib Q}_s) &=& 
\frac{f(E_{a'}({\mib k}+{\mib q}+{\mib Q}_s))-f(E_a({\mib k}))}
{\omega + E_a({\mib k}) -E_{a'}({\mib k}+{\mib q}+{\mib Q}_s) + i 0}. 
\end{eqnarray}

\begin{figure}
\begin{center}
\includegraphics[width=120mm]{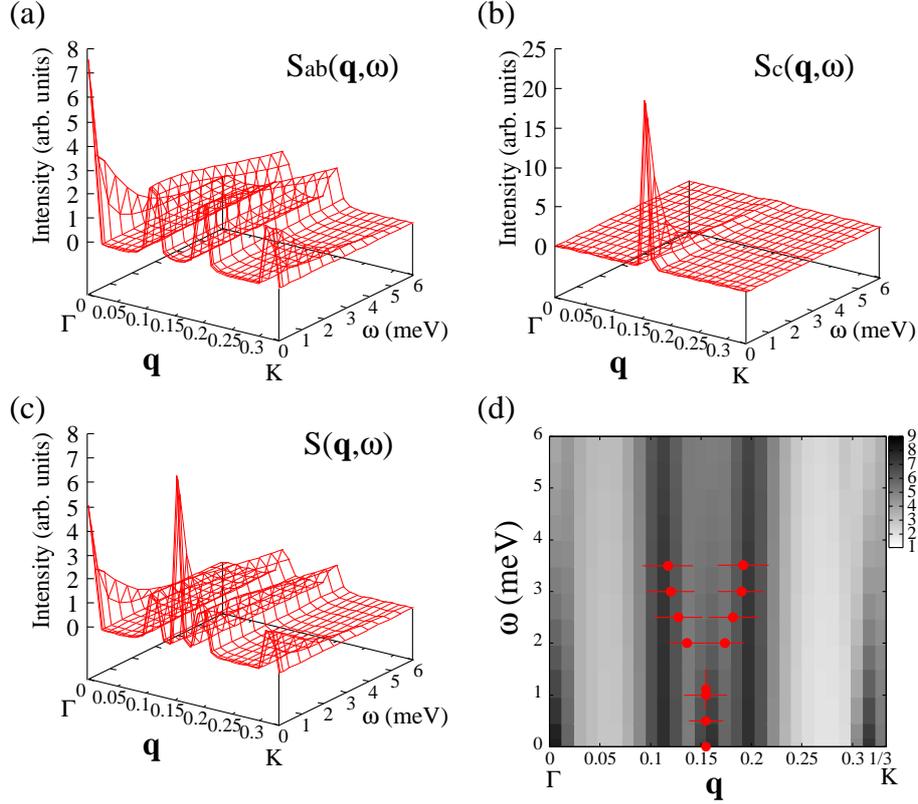}
\end{center}
\caption{(Color online) 
Calculated results of the dynamical spin correlation functions. 
(a) In-plane $S_{ab}({\mib q}, \omega)$ and (b) Out-of-plane $S_c({\mib q}, \omega)$, 
along the symmetry line $\Gamma$-K. (c) $S({\mib q},\omega)$ is the correlation function 
averaged over spin directions. 
(d) Comparison of $S({\mib q},\omega)$ with neutron scattering data. 
The dots with error bars are read from Ref.~\cite{Stock2010}. 
Gray-scale bar indicates intensity in a logarithmic scale.}
\label{Fig:6}
\end{figure}

Using the scattering function $\Lambda_{\nu,\zeta_3\zeta_4}({\mib q}, \omega; {\mib Q}_s)$, 
the dynamical spin correlation function is calculated by 
\begin{eqnarray}
S_{\mu\nu}({\mib q}, \omega) &=& \frac{2 \pi}{N} \sum_{\mib k} 
\sum_{\zeta_i} \sum_{s=0,\pm} 
[\Lambda_{\mu,\zeta_2\zeta_1}({\mib q}; {\mib Q}_s)]^* 
\Lambda_{\nu,\zeta_3\zeta_4}({\mib q}; {\mib Q}_s) 
u_{\zeta_4,a}({\mib k}) u_{\zeta_1,a}^*({\mib k}) \nonumber\\
&& \times u_{\zeta_2,a'}({\mib k}+{\mib q}+{\mib Q}_s) 
u_{\zeta_3,a'}^*({\mib k}+{\mib q}+{\mib Q}_s) 
f(E_a({\mib k})) [1-f(E_{a'}({\mib k}+{\mib q}+{\mib Q}_s))] \nonumber\\
&& \times \delta[\omega + E_a({\mib k}) - E_{a'}({\mib k}+{\mib q}+{\mib Q}_s)], 
\end{eqnarray}
which is represented diagrammatically in Fig.~\ref{Fig:5}(b). 
In-plane spin excitations are given by 
$S_{ab}({\mib q}, \omega) \equiv S_{xx}({\mib q}, \omega) = S_{yy}({\mib q}, \omega) $, 
while the out-of-plane one is $S_c({\mib q}, \omega) \equiv S_{zz}({\mib q}, \omega) $. 
When the spins of neutrons are not discriminated, the scattering cross section 
should be averaged over polarization directions, i.e., 
$S({\mib q},\omega) \equiv [2 S_{ab}({\mib q}, \omega) + S_c({\mib q}, \omega)]/3 $. 
Results of $S_{ab}({\mib q}, \omega)$, $S_c({\mib q}, \omega)$ and $S({\mib q}, \omega)$ 
are presented in Fig.~\ref{Fig:6}(a), (b) and (c), respectively, along the symmetry line $\Gamma-K$. 
A remarkable feature is that the in-plane excitations form peaks 
off the ordering vector ${\mib Q}$, not just at ${\mib Q}$. 
In addition, broadness along the energy axis is significantly different 
between the in-plane and out-of-plane modes. 

In the averaged $S({\mib q}, \omega)$, at low energies below 1.5 meV, 
a central main peak at ${\mib Q}$, which originates from out-of-plane spin excitations, 
is accompanied by weak satellites, which originate from in-plane spin excitations. 
Above 2 meV, while the out-of-plane spin excitations become weak, 
the in-plane spin excitations become relatively dominant. 
Calculation suggests the in-plane spin excitations yield 
also low-energy spectral weights near the $\Gamma$ and $K$ points. 
To compare with experiment, peak positions of neutron scattering spectra 
are overlaid on the calculated intensity map in Fig.~\ref{Fig:6}(d). 
Strong-intensity region around ${\mib Q}$ extends like a column along the excitation energy axis, 
whose shape is quite different from conventional V-shape for spin-wave modes in localized spin systems. 
If we interpret the experimental peak at ${\mib Q}$ below 1.5 meV as the out-of-plane spin excitation peak 
and the pair of the experimental peaks above 2.0 meV as the in-plane spin excitation peaks, 
then the calculated result seems not to contradict the experimental one. 
If realistic broadening is assumed, such a satellite structure may appear to be a shoulder or a tail 
around the main peak at ${\mib Q}$, yielding a broad spectrum around ${\mib Q}$. 

Thus, calculated $S({\mib q}, \omega)$ seems quite unusually different 
from that of conventional linear spin-wave excitations. 
On the other hand, this difference is not so surprising. 
In the weak-coupling MF theory, the spin response involves 
the degrees of freedom not only in spin rotation but also in spin norm. 
The norm degree of freedom does not vanish in the weak-coupling treatment, 
while it is neglected in localized spin approaches. 
Furthermore, the spin excitation modes are generally coupled with charge modes 
in incommensurate spiral ordering states even for much larger $U$~\cite{Cote1995,Brenig1996}. 
This coupling is completely neglected in localized spin approaches, 
where there are no degrees of freedom in the charge sector. 

\section{Concluding Remarks}
\label{Sc:4}

We have discussed the magnetic properties of a triangular lattice 
magnet NiGa$_2$S$_4$ within the spiral MF theory. 
If the Coulomb interaction among the Ni-$d$ states is weak $U \approx 2$ eV, 
then the most favorable magnetic state is the spiral ordering with ${\mib Q} \approx (0.15, 0.15)$, 
agreeing with neutron scattering. The Ni spin is calculated to be about 1.16 $\mu_B$ 
($|S| \approx 0.58$), similar to the experimental value $|S| \sim 0.51$. 
The significant reduction from $S=1$ is due to the Ni-S covalency rather than due to fluctuations. 

We consider the value of $U \approx 2$ eV is quite realistic for another reason: 
The value of $U$ to use in the MF theory should be regarded as already reduced 
from that of bare atomic Coulomb interaction to at most an effective width 
of the correlated bands, due to electron correlations~\cite{Kanamori1963}. 
For the present $d$-$p$ model, we can evaluate $U$ within the ladder approximation, 
as done for a single-band Hubbard model in Ref.~\cite{Chen1991}. 
According to our numerical calculation~\cite{Nomura2019}, $U$ is limited up to about 2 eV, 
since the effective width of the Ni-$d$ bands is about 2 eV. 

We also calculated the dynamical spin correlation function $S({\mib q}, \omega)$ within RPA, 
whose spectral structure quite differs from that of conventional spin-wave excitations. 
To confirm our results on the spectral properties of $S({\mib q}, \omega)$, 
one needs to resolve the in-plane and out-of-plane components carefully, 
using sufficiently polarized neutron scattering with high-purity single crystals. 

\begin{acknowledgments}
The authors are most grateful to Prof. K. Ishida, Prof. T. Sakai and Dr. I. Kawasaki 
for valuable conversations. 
\end{acknowledgments}

\appendix
\section{Consideration on the Heisenberg Spin Exchange Interactions in Magnetic Ordered States}

It is widely believed that low-energy properties of localized spin systems 
are well described by the Heisenberg spin Hamiltonian: 
\begin{equation}
H_S = - 2 \sum_{(i,j)} J(i,j) {\mib s}_i \cdot {\mib s}_j, 
\label{Eq:HA1}
\end{equation}
where ${\mib s}_i$ is spin operator at site $i$. 
Here let us consider that we calculate the Heisenberg exchange parameters $J(i,j)$ 
more microscopically from the underlying original Hubbard model with much larger $U$ than the hoppings. 
To do this, we usually treat the hopping terms in the Hubbard model as perturbation~\cite{Yosida}. 
$J(i,j)$'s are calculated by expanding perturbatively with respect to the hopping terms, 
where a pair of electrons are exchanged between sites $i$ and $j$ 
along all the exchange paths connecting the sites $i$ and $j$. 
If the system is in the non-magnetic state, each site on any of the exchange paths 
has up or down spin electron with even probability, in other words, is not polarized in spin. 
Let us define the nonmagnetic results of $J(i,j)$ calculated 
by this perturbative procedure, as $J_{\rm nonmag.}(i,j)$. 
It is naturally expected that $J_{\rm nonmag.}(i,j)$ should have the full symmetry of the lattice. 
Usually, the magnetic ground state, i.e., the most favorable spin ordered state is determined, 
using this $J_{\rm nonmag.}(i,j)$. 
However, we have to note here that, rigorously speaking, this $J_{\rm nonmag.}(i,j)$ 
is not appropriate to describe low energy properties in the magnetic ordered state, 
as we shall explain below.

In the magnetic ordered state, each of the localized spins is oriented to a favorable direction. 
Therefore the sites on the exchange paths are polarized in spin. 
As a result of the magnetic ordering, e.g., electrons with up spin 
cannot go through up-spin polarized sites on the exchange paths any more, 
whereas they could go through in the non-magnetic state since those sites were not polarized. 
Thus the situation is quite different from that in the non-magnetic state. 
If we repeat the above procedure to calculate $J(i,j)$ in the magnetic ordered state again, 
we will have different results of $J(i,j)$ ($\equiv J_{\rm mag.}(i,j)$), 
i.e. generally, $J_{\rm mag.}(i,j) \neq J_{\rm nonmag.}(i,j)$.

For another derivation, the exchange interactions $J(i,j)$ can be calculated by the relation~\cite{note} 
\begin{eqnarray}
\sum_{\sigma_n, n \neq i,j} \int d{\mib r}_1 \cdots d{\mib r}_N 
\psi^* ({\mib r}_1\sigma_1, \cdots, {\mib r}_i\sigma_i, \cdots, {\mib r}_j\sigma_j, \cdots, {\mib r}_N\sigma_N) \nonumber \\ 
\times P_{ij}^s [H - E_\psi]  P_{ij}^s 
\psi ({\mib r}_1\sigma_1, \cdots, {\mib r}_i\sigma'_i, \cdots, {\mib r}_j\sigma'_j, \cdots, {\mib r}_N\sigma_N) \nonumber \\
\equiv \langle \sigma_i\sigma_j | [-J(i,j)P_{ij}^s] | \sigma'_i\sigma'_j \rangle, 
\label{Eq:Jij}
\end{eqnarray}
where $H$ is the original Hamiltonian of the $N$-electron system, 
$\psi ({\mib r}_1\sigma_1, \cdots, {\mib r}_N\sigma_N)$ is the $N$-electron wave function normalized properly, 
and $P_{ij}^s$ is the permutation operator which exchanges 
spin states between the $i$-th and $j$-th electrons. 
$P_{ij}^s$ can be expressed effectively by using the spin operator~\cite{Yosida}: 
\begin{equation}
P_{ij}^s = \frac{1}{2}(1+4{\mib s}_i \cdot {\mib s}_j). 
\end{equation}
$H$ has the full symmetry of the system. 
In the nonmagnetic state, we have $J_{\rm nonmag.}(i,j)$ 
by using $\psi ({\mib r}_1\sigma_1, \cdots, {\mib r}_N\sigma_N)$ of the nonmagnetic state. 
In the magnetic ordered state, we have $J_{\rm mag.}(i,j)$ 
by using $\psi ({\mib r}_1\sigma_1, \cdots, {\mib r}_N\sigma_N)$ of the magnetic ordered state. 
Thus, generally, $J_{\rm mag.}(i,j) \neq J_{\rm nonmag.}(i,j)$, 
since the wave function $\psi ({\mib r}_1\sigma_1, \cdots, {\mib r}_N\sigma_N)$ is different 
between the nonmagnetic and magnetic ordered states.

$J_{\rm mag.}(i,j)$ will depend on the ordered spin configuration. 
As naturally expected, low-energy properties in the magnetic ordered state, such as spin waves, 
should be described not by $J_{\rm nonmag.}(i,j)$ but by $J_{\rm mag.}(i,j)$. 
Here note that $J_{\rm mag.}(i,j)$'s generally do not always have the full symmetry of the lattice, 
particularly if the magnetic ordering has lower symmetry than the lattice. 
To our knowledge, there is no study estimating how much $J_{\rm mag.}(i,j)$ deviates 
from $J_{\rm nonmag.}(i,j)$, although it seems an interesting issue.

Now we turn our attention to the case of the spiral ordering in NiGa$_2$S$_4$. 
$J_{n(n')}$ calculated in \S~\ref{Sc:3.3} corresponds 
to $J_{\rm mag.}(n(n'),0)$, not to $J_{\rm nonmag.}(n(n'),0)$. 
In eq.~(\ref{Eq:Jni}), all the virtual exchange (hopping) processes between the sites $i$ and $n$ are included, 
although spin polarization at each site on the exchange paths is treated only as a static MF potential. 
As understood from the above consideration, possibility of each exchange process 
depends on the hopping direction relative to the direction of the ordering vector ${\mib Q}$. 
Thus we need to classify the Heisenberg exchange parameters into $J_{n}$ and $J_{n'}$, 
since the spiral ordering breaks the $C_3$ rotational symmetry.


\begin{thebibliography}{99}
\bibitem{note}
Note that the relation eq.~(\ref{Eq:Jij}) as well as eq.~(\ref{Eq:Jni}) 
is obtained from the total-energy variation induced adiabatically 
by spin exchange or rotation of the $i$-th and $j$-th electrons. 
A similar formula to eq.~(\ref{Eq:Jij}) is seen in \S 4.4 of Ref.~\cite{Yosida}.
\end{thebibliography}
\end{document}